\documentclass[nofootinbib,aps,11pt]{revtex4-1}
\usepackage{graphicx}
\usepackage{subfigure} 
\usepackage{hyperref}
\usepackage{cancel}
\usepackage{amssymb}
\usepackage{textcomp}
\usepackage{amsmath}
\usepackage{bm}
\usepackage{times}
\usepackage{epsfig}
\usepackage{color}
\usepackage{mathrsfs}
\newcommand{\rh}{{\rm rh}}
\newcommand{\fo}{{\rm fo}}
\newcommand{\eq}{{\rm eq}}
\newcommand{\eff}{{\rm eff}}
\newcommand{\GeV}{{\rm GeV}}
\newcommand{\MeV}{{\rm MeV}}
\newcommand{\cm}{{\rm cm}}

\begin{document}
\title{\Large Dark photon portal dark matter with low-temperature reheating }
	\bigskip
\author{Zhi-Long Han$^1$}
\email{sps\_hanzl@ujn.edu.cn}
\author{Honglei Li$^1$}
\email{sps\_lihl@ujn.edu.cn}
\author{Ang Liu$^2$}
\email{AL@jnxy.edu.cn}
\author{Lei Wu$^{3}$}
\email{leiwu@njnu.edu.cn}
\author{Cai-Xia Yang$^1$}
\email{ycxmumu@qq.com}
\affiliation{$^1$School of Physics and Technology, University of Jinan, Jinan, Shandong 250022, China}
\affiliation{$^2$School of Physics and Electronic Engineering, Jining University, Shandong 273155, China}
\affiliation{$^3$Department of Physics and Institute of Theoretical Physics, Nanjing Normal University, Nanjing 210023, China}	
	\date{\today}
	
\begin{abstract} 
The dark photon $A'$ is widely considered as the mediator of dark matter $\chi$. In the  conventional non-resonance benchmark scenario with $m_{A'}/m_\chi=3$,  the kinetic mixing $\epsilon$  required to match the observed dark matter relic density is typically ruled out by the combined constraints from the direct detection, indirect detection, collider, and other relevant experiments. However, if a delayed decay of the inflaton creates the low-temperature reheating, the additional entropy production dilutes the dark matter relic density. As a result, a significantly smaller $\epsilon$ becomes sufficient to match the observation, which allows the dark matter to escape the present multi‑experimental bounds. In this paper, we investigate the dark matter production under the influence of a low reheating temperature $T_{\rh}$ within the dark photon $A^\prime$ portal framework, where  $A^\prime$  mediates the interaction between dark matter $\chi$  and the SM particles.  We  systematically explore  the viable and promising parameter space for  complex scalar, Dirac, and Majorana fermion dark matter under the combined experimental constraints, and compare the distinctions among these three scenarios.
\end{abstract}

	\maketitle

\section{Introduction}

The cosmological and astrophysical observations have provided compelling evidence for the existence of dark matter (DM) \cite{Cirelli:2024ssz}. However, the nature of dark matter remains an open question. DM could be composed of particles \cite{Bertone:2004pz}. The prevailing view is that it cannot be accounted for by any Standard Model (SM) particle. Hence, it is necessary to extend the SM with a new particle that serves as the viable dark matter candidate. A simple extension is to introduce a vector field called the dark photon $A^\prime$, which is identified as the gauge boson of an additional Abelian $U(1)^\prime$ gauge symmetry, and has kinetic mixing $\epsilon$ with the photon field or the hypercharge field \cite{Caputo:2026pdw,Fabbrichesi:2020wbt}. A dark photon below MeV could serve as the dark matter candidate \cite {Nelson:2011sf,Arias:2012az,Aboubrahim:2021ycj,Feng:2024nkh}, but the critical density and cosmic microwave background (CMB) observations impose extremely stringent constraints on this scenario \cite{Fabbrichesi:2020wbt}.

A more appealing scenario is that $A^\prime$ acts as a mediator connecting the dark matter $\chi$ and the SM \cite{He:1991qd, Chang:2018rso, Hambye:2019dwd, Filippi:2020kii,Emken:2024nox,Mishra:2025juk,Cheek:2025nul,Liang:2026rqp},  meanwhile $\chi$ freezes out from the SM thermal bath as a weakly interacting massive particle (WIMP) \cite{Arcadi:2017kky,Roszkowski:2017nbc}. Such new physics models often face the stringent constraints from the DM direct detection \cite{DarkSide-50:2023fcw,CRESST:2019jnq,CRESST:2024cpr,XENON:2025vwd,PandaX:2024qfu,LZ:2024zvo,DAMIC-M:2025luv,DarkSide:2022knj,PandaX:2022xqx}, indirect detection \cite{Fermi-LAT:2015att,Fermi-LAT:2016uux,HESS:2016mib,CTA:2020qlo,HESS:2022ygk}, and accelerator searches for $A^\prime$ \cite{Bjorken:2009mm,Baumgart:2009tn,APEX:2011dww,Curtin:2014cca,Ilten:2016tkc,LHCb:2017trq,BESIII:2017fwv,Hearty:2022wij,PADME:2026any,LDMX:2026xum,Zhang:2026cpk,Chen:2026xue}. The viable $\epsilon$ allowed by these combined constraints is very small, and its specific magnitude  is related to the properties of dark particles  as well as the mass scope. Under the condition of satisfying the observed dark matter relic density \cite{Planck:2018vyg}, the secluded dark matter scenario with $m_\chi>m_{A^\prime}$ can readily avoid  these combined constraints, since the production of $\chi$, namely,  $\chi\chi\to A^\prime A^\prime$, does not depend on $\epsilon$ \cite{Pospelov:2007mp,Mohapatra:2019ysk,Das:2026buc}. For the same but kinematically suppressed annihilation process, the forbidden dark matter with $1\lesssim m_{A^\prime}/m_\chi \lesssim 2$ is also viable \cite{Griest:1990kh,DAgnolo:2015ujb,Cline:2017tka,Fitzpatrick:2020vba,Fitzpatrick:2021cij}. In the light $m_\chi$ regime with $m_{A^\prime}/m_\chi \gtrsim 2$, the resonance scenario is allowed by the combined constraints owing to the suppression of $\epsilon$, whereas the non‑resonant case, e.g., $m_{A^\prime}/m_\chi=3$, remains very challenging \cite{Nath:2021uqb,Krnjaic:2025noj,Wang:2025clh,Alonso-Gonzalez:2025xqg}. Naturally, alternative dark matter production mechanisms may alleviate this issue, for example, the coannihilation \cite{Griest:1990kh} and coscattering \cite{DAgnolo:2017dbv} mechanisms  in the inelastic DM scenarios \cite{Filimonova:2022pkj,Zhang:2024sox,Wang:2025fya}, as well as the low-temperature
reheating scenarios \cite{Giudice:2000ex,Fornengo:2002db,Gelmini:2006pw,Drees:2006vh,Roszkowski:2014lga,Bernal:2022wck,Haque:2023yra,Bernal:2023ura,Chowdhuryand:2024uvi,Boddy:2024vgt,Bernal:2024yhu,Barman:2024lxy,Bernal:2024ndy,Belanger:2024yoj,Barman:2024nhr,Barman:2024tjt,Borah:2025ema,Khan:2025kuh,C:2026bqd,Bertou:2026osq}.

In this paper, we focus on the low-temperature reheating mechanism.  We take into account that the inflaton $\phi$ decays into the SM during the reheating period, thereby increasing the entropy of the universe \cite{Allahverdi:2020bys,Batell:2024dsi}. This entropy injection modifies the reheating temperature $T_{\rh}$, which marks the onset of radiation domination. A longer lifetime of $\phi$ results in a smaller $T_{\rh}$, which is typically above a few MeV \cite{Sarkar:1995dd, Kawasaki:2000en, Hannestad:2004px, DeBernardis:2008zz, deSalas:2015glj}. In such a low $T_{\rh}$ scenario, the coupling needed to match the observed relic abundance is smaller than that of the conventional WIMP paradigm by several orders of magnitude \cite{Gelmini:2006pw,Bernal:2022wck,Bernal:2024yhu}. Motivated by this, we investigate the production of the complex scalar, Dirac, and Majorana fermion dark matter $\chi$ under a low $T_{\rh}$ environment, and derive the viable parameter space under the combined experimental constraints  within a simplified dark photon portal framework  where only $A^\prime$ and $\chi$ are introduced beyond the SM.

The structure of this paper is organized as follows. In Section \ref{SEC:TM}, we provide a brief introduction to the theoretical  model.  Section \ref{SEC:LR} is devoted to the calculation of dark matter production under the condition of low-temperature reheating. The various relevant phenomenological constraints are considered in Section~\ref{SEC:PC}. In Section \ref{SEC:CR}, we identify the viable parameter regions that could be probed by future experiments in light of the combined constraints. Finally, we summarize the results in Section \ref{SEC:CL}.

\section{The model}\label{SEC:TM}

This work is based on a simplified  model, in which the dark photon $A^\prime$ mediates the interactions between the singlet dark matter $\chi$ and  SM fermions.  The relevant Lagrangian can be written as 
\begin{eqnarray}\label{Eqn:lg}
	\mathcal{L}_{\rm int}=-  A^\prime_\mu(  g_\chi J_\chi^\mu +   \epsilon e J_{\rm EM}^\mu  ),
\end{eqnarray}
where $g_\chi = \sqrt{4\pi \alpha_\chi}$  is the coupling between dark matter and the dark photon, whose corresponding  interaction  originates from the covariant derivative  term of the dark photon field.  $\epsilon$ is the kinetic mixing parameter that governs interactions with the electromagnetic current $J_{\rm EM}$ \cite{Fabbrichesi:2020wbt,Caputo:2026pdw}, and $e$ is  the electromagnetic
coupling constant. Depending on the nature of $\chi$, $J_\chi$ can be expanded as \cite{Krnjaic:2025noj}
\begin{eqnarray}
	J_D^\mu = 
	\begin{cases}
		~~i\chi^* \partial^\mu \chi + c.c. & \text{Complex Scalar},  
		\\
		~~\overline \chi \gamma^\mu \chi  & \text{Dirac Fermion},
		\\
	    ~~\frac{1}{2}\overline \chi \gamma^\mu\gamma^5 \chi  & \text{Majorana Fermion}.
	\end{cases}
\end{eqnarray}
Meanwhile, the electromagnetic current can be explicitly written as
\begin{eqnarray}
	J_{\rm EM}^\mu = 
	\begin{cases}
		~~-\bar{l}\gamma^\mu l & l=e, \mu, \tau,  
		\\
		~~\frac{2}{3}\bar{q}\gamma^\mu q  & q=u, c, t,
		\\
		~~-\frac{1}{3}\bar{q}\gamma^\mu q  & q=d, s, b.
	\end{cases}
\end{eqnarray}

In this model, $A^\prime$ and $\chi$ have masses $m_{A^\prime}$ and $m_\chi$, respectively. We remain agnostic about the origin of these masses. Notably, in a more complete framework, the parameter $\epsilon$ represents the mixing between the dark photon field and the hypercharge field. After the  electroweak symmetry breaking, dark matter couples not only to $A^\prime$ but also to the $Z$ boson, and there additionally exists neutral currents in the SM part \cite{Babu:1997st,Curtin:2014cca,Cirelli:2016rnw,Filimonova:2022pkj}. Under the present collider sensitive region $m_{A^\prime}/m_Z\ll1$, the contributions of neutral currents are suppressed by $m_{A^\prime}/m_Z$, and thus are far smaller than the charged current in Equation \eqref{Eqn:lg}. Moreover, we devote particular attention to the contribution of the $A^\prime$ portal dark matter, and considering the coupling of the Higgs portal term  $(H^\dagger H)(\chi^\dagger\chi)$ to be zero for the scalar dark matter.

\section{Low-temperature reheating and DM relic density}\label{SEC:LR}

The conventional WIMP scenario assumes that the reheating temperature $T_{\rh}$ is  much greater than the decoupling temperature $T_{\fo}\sim m_\chi/20$, so that the dark matter relic density is not affected. In contrast, we consider a reheating epoch in which the inflaton  $\phi$ delayed decays into the SM radiation with a decay width of $\Gamma_\phi$. A sufficiently small $\Gamma_\phi$ will postpone the reheating temperature to a low scale, e.g., $T_{\rh}\lesssim T_{\fo}$, thereby significantly affecting the generation of dark matter.  The value of $T_{\rh}$ is rather loosely constrained, as long as it satisfies $T_{\rh}>T_{\rm BBN}\simeq4$ MeV \cite{Sarkar:1995dd, Kawasaki:2000en, Hannestad:2004px, DeBernardis:2008zz, deSalas:2015glj}, which does not affect the Big Bang Nucleosynthesis (BBN) observations. The cosmological evolution of the relevant inflaton energy density $\rho_\phi$,  SM entropy density $s$, and number density $n_\chi$ of DM,  could be described by the following Boltzmann equations \cite{Gelmini:2006pw,Belanger:2024yoj}
\begin{eqnarray}\label{Eqn:be}  
\frac{dz_\phi}{da} &=& -\frac{\Gamma_\phi}{Ha}z_\phi,\\ \nonumber
\frac{dz_s}{da} &=& \frac{4\Gamma_\phi s^{1/3}}{3HT}z_\phi,\\ \nonumber
\frac{dz_\chi}{da} &=& -\frac{1}{Ha^4}\langle\sigma v \rangle(z_\chi^2-({z_\chi^\eq})^2),
\end{eqnarray}
where the parameters are determined as $z_\phi \equiv \rho_\phi \times a^3$, $z_s \equiv s^{4/3} \times a^4$,  $z_\chi= n_\chi \times a^3$, and  $z_\chi^\eq= n_\chi^\eq \times a^3$, with the cosmic scale factor $a$. 

The entropy density is defined as a function of the SM temperature $T$
\begin{eqnarray}
	s(T) = \frac{2\pi^2}{45} h_{\eff}(T) T^3,
\end{eqnarray}
with $h_{\eff}(T)$ the number of relativistic degrees of freedom for the entropy density. The Hubble expansion rate  satisfies
\begin{eqnarray}
	H^2 = \frac{8\pi}{3} \frac{\rho_\phi+\rho_R}{m_{pl}^2},
\end{eqnarray}
with the Planck mass  $m_{pl}\simeq 1.2 \times 10^{19}$~GeV, and the SM energy density $\rho_R$ is given by
\begin{equation}
	\rho_R(T) = \frac{\pi^2}{30} g_{\eff}(T) T^4,
\end{equation}
where $g_{\eff}(T)$ is the number of relativistic degrees of freedom for the SM energy density. 

In the evolution equation of DM, $\langle\sigma v\rangle$ is the thermally averaged DM annihilation cross section, which mainly involves the processes mediated by the dark photon $A^\prime$. $n_\chi^{\eq}$ is the DM number density at equilibrium, which is expressed as
\begin{equation}
	n_\chi^{\eq}(T) = 2 \left(\frac{m_\chi T}{2\pi}\right)^{3/2} e^{-m_\chi/T},
\end{equation}
in the non-relativistic limit.

 \begin{figure}
 	\begin{center}
 		\includegraphics[width=0.45\linewidth]{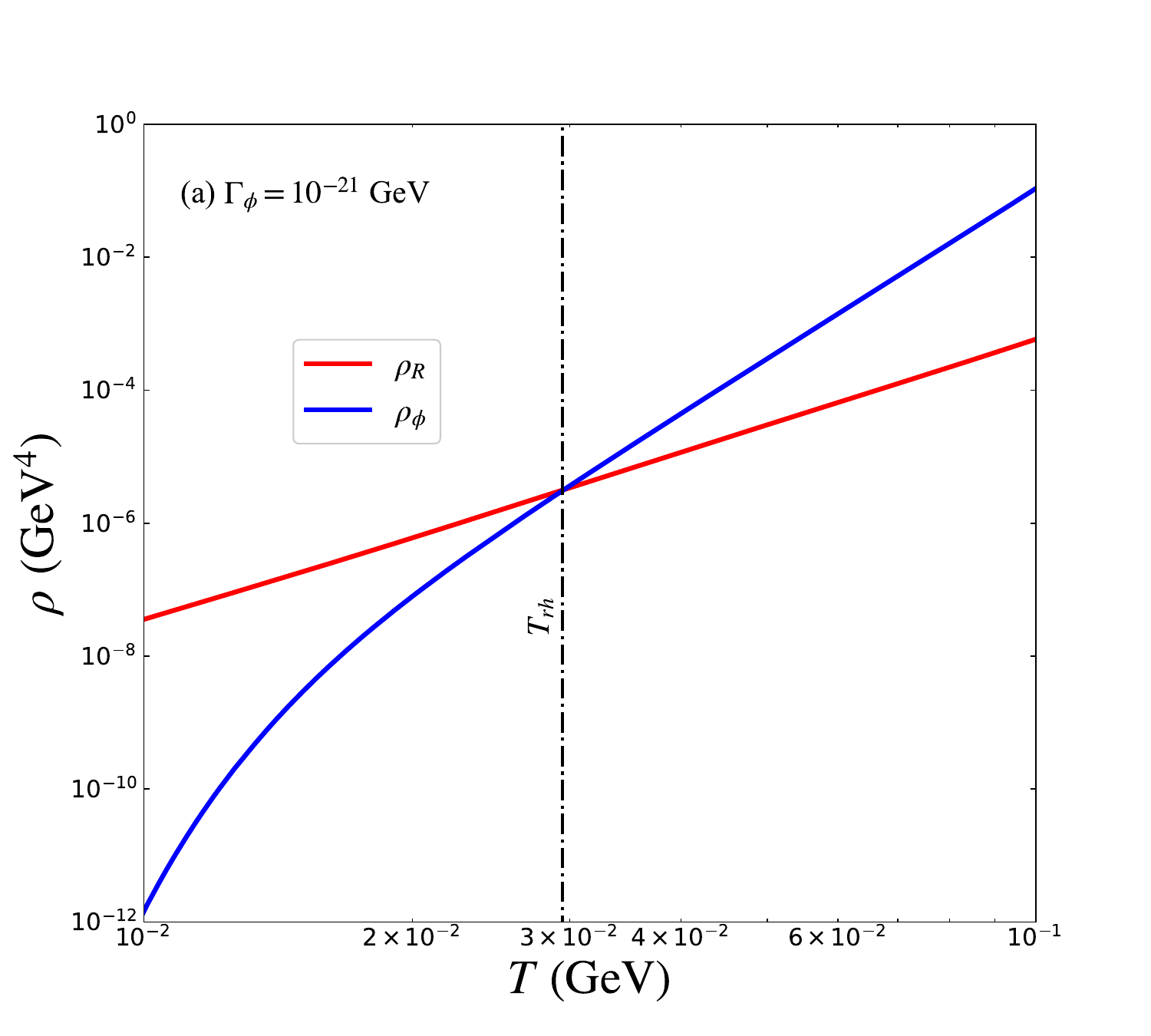}
 		\includegraphics[width=0.45\linewidth]{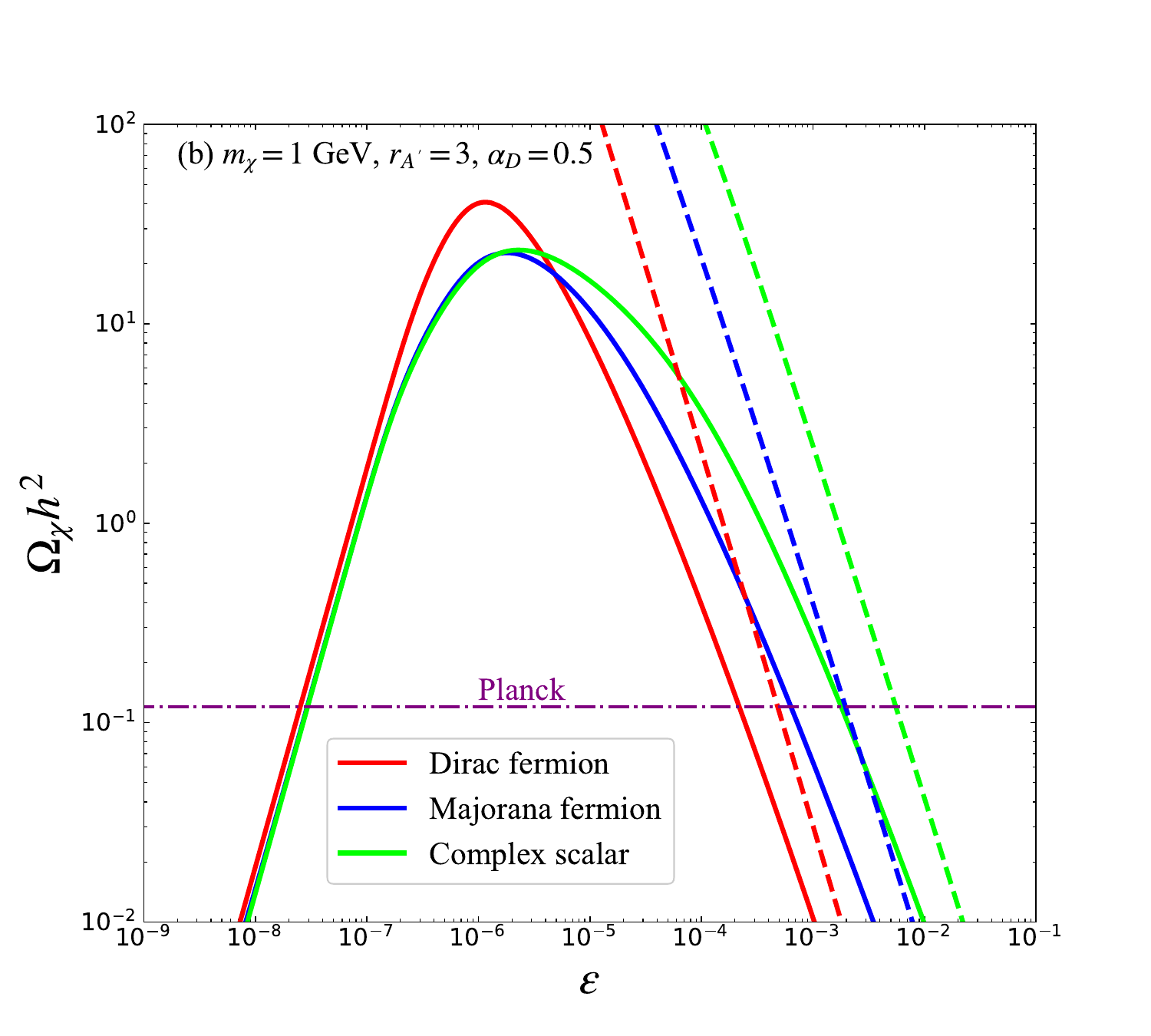}
 	\end{center}
 	\caption{Panel (a): The temperature dependence of $\rho_\phi$ and $\rho_R$. The red and blue solid lines represent $\rho_\phi$ and $\rho_R$, respectively, and the vertical dot-dashed line corresponds to $T_{\rh}$. Panel (b): The relic density of dark matter $\Omega_\chi h^2$ as a function of $\epsilon$. The red, blue, and green curves represent the Dirac, Majorana, and complex scalar dark matter, respectively. Meanwhile the solid and dashed lines denote the scenarios with low $T_{\rh}$ effects and conventional WIMP dark matter, respectively. The purple horizontal dot-dashed line indicates the observed relic density of dark matter, i.e., $\Omega_\chi h^2=0.12$ \cite{Planck:2018vyg}.
 	}
 	\label{FIG:fig1}
 \end{figure}
The solution of the Boltzmann Equations~\eqref{Eqn:be} relies on micrOMEGAs \cite{Alguero:2023zol,Belanger:2026asz}. The improved precision in the calculations of $h_{\eff}(T)$ and $g_{\eff}(T)$ \footnote{Within the range $\mathcal{O}(1)~\MeV \lesssim T_{\rh}\lesssim \mathcal{O}(100)~\MeV$ of interest in this work, the numerical values of $h_{\eff}$ and $g_{\eff}$ show little variation, staying near 11.}  prevents an $\mathcal{O}(10\%)$ change in the relic density of the GeV scale dark matter \cite{Drees:2015exa}, which decouples  around the QCD phase transition
temperature $T_{\rm QCD}\sim0.2$ GeV. We show the results for the benchmark point $(\Gamma_\phi=10^{-21}~ \GeV, m_\chi=1~\GeV, r_{A^\prime}=m_{A^\prime}/m_\chi=3, \alpha_\chi=0.5)$ in  Figure~\ref{FIG:fig1}. Panel (a) of  Figure~\ref{FIG:fig1} displays the evolution of $\rho_\phi$ and $\rho_R$ with the temperature $T$. As the universe cools, $\rho_\phi$ starts to fall below $\rho_R$ at $T_\rh=T\simeq29.4$ MeV, resulting the end of the reheating epoch. Then the universe becomes dominated by SM radiation. Under the condition $H(T_\rh) = \Gamma_\phi$, $T_\rh$ could be approximately expressed as \cite{Belanger:2024yoj}
\begin{equation}\label{Eqn:trh}
	T_\rh^2 = \frac{3}{2\pi} \sqrt{\frac{5}{\pi g_{\eff}(T_{\rh})}}m_{pl}\Gamma_\phi.
\end{equation}
 
Regarding the dark matter production, for the conventional WIMP scenario without low $T_{\rh}$ effects, as indicated by the dashed curves in panel (b) of Figure~\ref{FIG:fig1}, the $\epsilon$ compatible with the observed DM relic density is relatively large. The specific values vary with the nature of DM, but generally lie in the range of $\mathcal{O}(10^{-4})\sim\mathcal{O}(10^{-2})$. Such magnitude poses a severe challenge to the dark matter (in)direct detection and collider constraints, which could reach  $\epsilon\sim\mathcal{O}(10^{-5})$. A detailed discussion is deferred to Section \ref{SEC:CR}. When the $T_{\rh}$ effects are included, depending on whether the DM reaches thermal equilibrium, there emerge two $\epsilon$ solutions satisfying the observed density, which corresponds to the freeze-out and freeze-in mechanisms, respectively. The extremely small freeze-in solution could be as low as $\epsilon\sim\mathcal{O}(10^{-8})$. Meanwhile, even the larger freeze-out solution is smaller than the corresponding WIMP result, due to the dilution of the relic density by entropy injection. Owing to this, dark matter in the non‑resonant regime can evade exclusion from various experimental constraints.

\section{Phenomenological constraints}\label{SEC:PC}

\subsection{Direct detection} \label{SUBSEC:DD}

At present, the nuclear recoil experiments DarkSide-50 \cite{DarkSide-50:2023fcw}, CRESST-III \cite{CRESST:2019jnq, CRESST:2024cpr},  XENONnT \cite{XENON:2025vwd}, PandaX-4T \cite{PandaX:2024qfu}, LZ \cite{LZ:2024zvo},  etc., impose stringent constraints on the spin-independent $\chi-n$ scattering cross section with $m_\chi$ above $\mathcal{O}(10)$ MeV.  For $m_\chi$ below GeV, electron recoil is also an effective detection pathway.  The related experiments DAMIC-M \cite{DAMIC-M:2025luv}, DarkSide-50 \cite{DarkSide:2022knj}, PandaX-4T \cite{PandaX:2022xqx}, etc., also set stringent limits on the $\chi-e$ scattering cross section.   The future nuclear recoil experiments  DarkSide-LowMass \cite{GlobalArgonDarkMatter:2022ppc}, SuperCDMS \cite{SuperCDMS:2016wui}, and LZ \cite{LZ:2015kxe}, as well as the electron recoil experiment BRN  \cite{Essig:2022dfa} will further extend the exploration scope.

Turning to our specific model, the scattering cross sections for the complex scalar and Dirac fermion DM scenarios are numerically calculated through \cite{Essig:2011nj}
\begin{equation}\label{Eqn:dd-e}
\sigma_e = \frac{16\pi \alpha \alpha_\chi \epsilon^2 \mu_{\chi e}^2 }{ m_{A^\prime}^4},
\end{equation}
where $\alpha$ is the  fine-structure constant, and $\mu_{\chi e}$ is the reduced mass of DM $\chi$ and electron $e$. For the $\chi-n$ scattering cross section $\sigma_n$, one needs to replace $\mu_{\chi e}$ with $\mu_{\chi n}$. In the heavy mediator scenario with the form factor $F_{\rm DM}=1$, using the conversion relation between $\sigma_e$ and $\sigma_n$, we obtain the most stringent current combined constraints from electron recoil and nuclear recoil experiments in panel (a) of Figure~\ref{FIG:fig2} and Figure~\ref{FIG:fig4}, shown as the orange solid line. This line provides the strongest limit around 10 GeV, excluding the parameter space with $\sigma_e$ above $\mathcal{O}(10^{-53})~\rm cm^2$. The future constraints, indicated by the orange dashed line, show that the lower limit of the sensitivity  drops from $\mathcal{O}(10^{-43})~\rm cm^2$ to $\mathcal{O}(10^{-55})~\rm cm^2$ as $m_\chi$ increases from 0.01 GeV to 10 GeV.

However, in the Majorana dark matter scenario, the  $\chi-e$ and $\chi-n$ scattering cross sections depend on the velocity $v$ of dark matter \cite{Berlin:2018bsc}. The corresponding $F_{\rm DM}$ is proportional to $q^2$. The stringent electron and nuclear recoil constraints are obtained by fixing $F_{\rm DM}=1$ with the heavy mediator, which are not applicable in this scenario. Since the axial-vector structure in the Majorana  scenario cannot generate spin-dependent operators, we cannot use the spin-dependent nuclear recoil constraints either \cite{Hooper:2014fda}.

\subsection{Indirect detection} \label{SUBSEC:ID}

The goal of DM indirect detection experiments is to detect DM via the SM particles generated by its annihilation. We take the electron final state as an example, not only because it has a significant contribution in this model, but also because this process is always kinematically allowed in the mass range MeV-GeV that we consider. For $m_\chi$ above GeV, the most stringent upper limit from current experiments on  $\langle \sigma v\rangle$  is summarized in Refs.~\cite{Leane:2018kjk,Dutta:2022wdi}, which are the convolutions of the bounds from AMS positron \cite{AMS:2014xys,AMS:2019rhg}, Fermi-LAT dwarfs \cite{Fermi-LAT:2016uux} and
H.E.S.S. GC observations \cite{HESS:2016mib,HESS:2022ygk}. For the lighter dark matter, in particular the $s$-wave case, CMB temperature anisotropies place strong constraints on the energy injection \cite{Cirelli:2023tnx,Wang:2025jhy,Cirelli:2025rky}, with the Planck data requiring the total $\langle \sigma v\rangle$ to satisfy \cite{Planck:2018vyg}
\begin{equation}
\langle \sigma v\rangle \lesssim  2 \times 10^{-26}~{\rm cm^3/s} \left(\frac{0.4}{f_{\rm \eff}}\right) \left(\frac{m_\chi}{30~\GeV}\right),
\end{equation}
where we optimistically take the ionization efficiency factor $f_{\rm eff}=0.4$ \cite{Slatyer:2015jla}. The combined constraints reach their maximum exclusion  for $\langle \sigma v\rangle\gtrsim\mathcal{O}(10^{-30})~\rm cm^3/s$ at $m_\chi=10^{-2}$ GeV, as shown by the gray shaded regions in panel (b) of Figure~\ref{FIG:fig2} and Figure~\ref{FIG:fig3}, and in  panel (a) of Figure~\ref{FIG:fig4}.

For the future MeV telescopes AMEGO \cite{AMEGO:2019gny,Kierans:2020otl,Caputo:2022xpx},
E-ASTROGAM \cite{e-ASTROGAM:2016bph,e-ASTROGAM:2017pxr} and MAST \cite{Dzhatdoev:2019kay}, the projected sensitivities within $m_\chi\gtrsim0.3$ GeV  have been derived \cite{Cirelli:2025qxx}, shown as the black dashed curve, which is about two orders of magnitude below the existing experimental bound.  Within the lighter mass range $10^{-2}~\GeV\lesssim m_\chi \lesssim0.3~\GeV$ investigated in our study, the projected sensitivity of the  upcoming MeV telescope COSI \cite{Tomsick:2019wvo,Beechert:2022phz} is derived \cite{Cirelli:2025rky}, which has $\langle \sigma v\rangle\sim \mathcal{O}(10^{-29})~\rm cm^3/s$, the sensitivity is of a similar order of magnitude for lighter $\mathcal{O}(1)$ MeV dark matter \cite{Saha:2025wgg}. This magnitude cannot probe the complex scalar or Majorana fermion dark matter, and in the Dirac case it is already excluded by the BBN constraint. Therefore, we do not show this part in the figures. In order to obtain more accurate results, we use micrOMEGAs \cite{Alguero:2023zol,Belanger:2026asz}  to compute  the specific $\langle \sigma v\rangle$.

\subsection{Supernova cooling and electroweak precision tests} \label{SUBSEC:CE}

The invisible dark particles produced in the proto-neutron star  can easily escape and carry away energy due  to their feeble interactions, thereby accelerating the cooling of the supernova, which is constrained by relevant observations. In the scenario where the dark photon decays mainly into dark matter, the deriving supernova SN1987A cooling constraints \cite{Chang:2018rso} rule out $m_{A^\prime}\lesssim\mathcal{O}(100)$ MeV with $\mathcal{O}(10^{-10})\lesssim \epsilon \lesssim\mathcal{O}(10^{-7})$ \footnote{The constrained parameter space  is not significantly different in the model without dark matter \cite{Chang:2016ntp, Caputo:2025avc}. }. Upon our verification,  such excluded parameter space almost all fall into the region already excluded by BBN with $T_{\rh}<4$ MeV. Hence, we have not shown this constraint in this paper.

In addition, when $\epsilon$  originates from the mixing with the hypercharge field, it affects the electroweak precision observables of $Z$ boson. Within our considered range of $m_{A^\prime}/m_Z\ll1$, the current LHC impose a $95\%$ CL limit $\epsilon\lesssim0.019$ \cite{Curtin:2014cca}. The future ILC/GigaZ  will be able to test $\epsilon$ above $3.5\times10^{-3}$ \cite{Curtin:2014cca}. In the complex scalar and Dirac fermion DM  scenarios, this type  of constraint is far less stringent than the direct detection limits. Therefore, we do not show it in the relevant figures. However, for the Majorana case  where the direct detection constraints are absent,  it is meaningful to discuss this constraint, which corresponds to the purple curves in Figure \ref{FIG:fig4}.

\subsection{Collider signals} \label{SUBSEC:CS}

In this work, since we set $m_{A^\prime}=3 m_\chi$ and $\alpha_\chi=0.5$, the invisible decay of $A^\prime$ is the dominant channel, i.e., $\text{Br}(A^\prime\to\bar{\chi}\chi)\simeq1$. In order to be consistent with our analysis, we take  the limits on invisible decay of $A^\prime$  from the  NA64 experiment \cite{NA64:2016oww,NA64:2017vtt,Banerjee:2019pds} through the searching for missing energy events in the electron-nucleus scattering,  and from  BABAR \cite {BaBar:2017tiz}, BESIII \cite {Zhang:2019wnz} and LEP \cite{DELPHI:2003dlq,DELPHI:2008uka,Fox:2011fx} via the monophoton searches. Their combined results on dark matter are shown in the gray shaded region in panel (c) of Figure~\ref{FIG:fig2} and \ref{FIG:fig3}, and in  panel (b) of Figure~\ref{FIG:fig4}, where  $y=\epsilon^2 \alpha_\chi (m_{\chi}/m_{A^\prime})^4$. These current constraints exclude the parameter space with $y\gtrsim\mathcal{O}(10^{-10})$ (corresponding to $\epsilon\gtrsim \mathcal{O}(10^{-4})$) when $m_\chi$ is less than  10 GeV.

The future STCF \cite{Charm-TauFactory:2013cnj,Zhang:2019wnz} and DarkSHINE \cite{Li:2025nny,Yang:2025yvl} experiments will probe the invisible dark photon via the monophoton signature and the missing energy/momentum signature, respectively. The resulting sensitivity is about two orders of magnitude below the present limits.

\section{Combined results}\label{SEC:CR}

Combining the phenomenological constraints discussed above, we respectively present in Figure~\ref{FIG:fig2}, Figure~\ref{FIG:fig3}, and Figure~\ref{FIG:fig4} the viable parameter space for the  complex scalar, Dirac, and Majorana fermion dark matter with low reheating temperatures. We focus on the conventional non-resonance benchmark $r_{A^\prime}=3$ and $\alpha_\chi=0.5$ throughout this work.

\subsection{Complex scalar scenario} \label{SUBSEC:CSS}

\begin{figure}[h!]
	\begin{center}
		\includegraphics[width=0.45\linewidth]{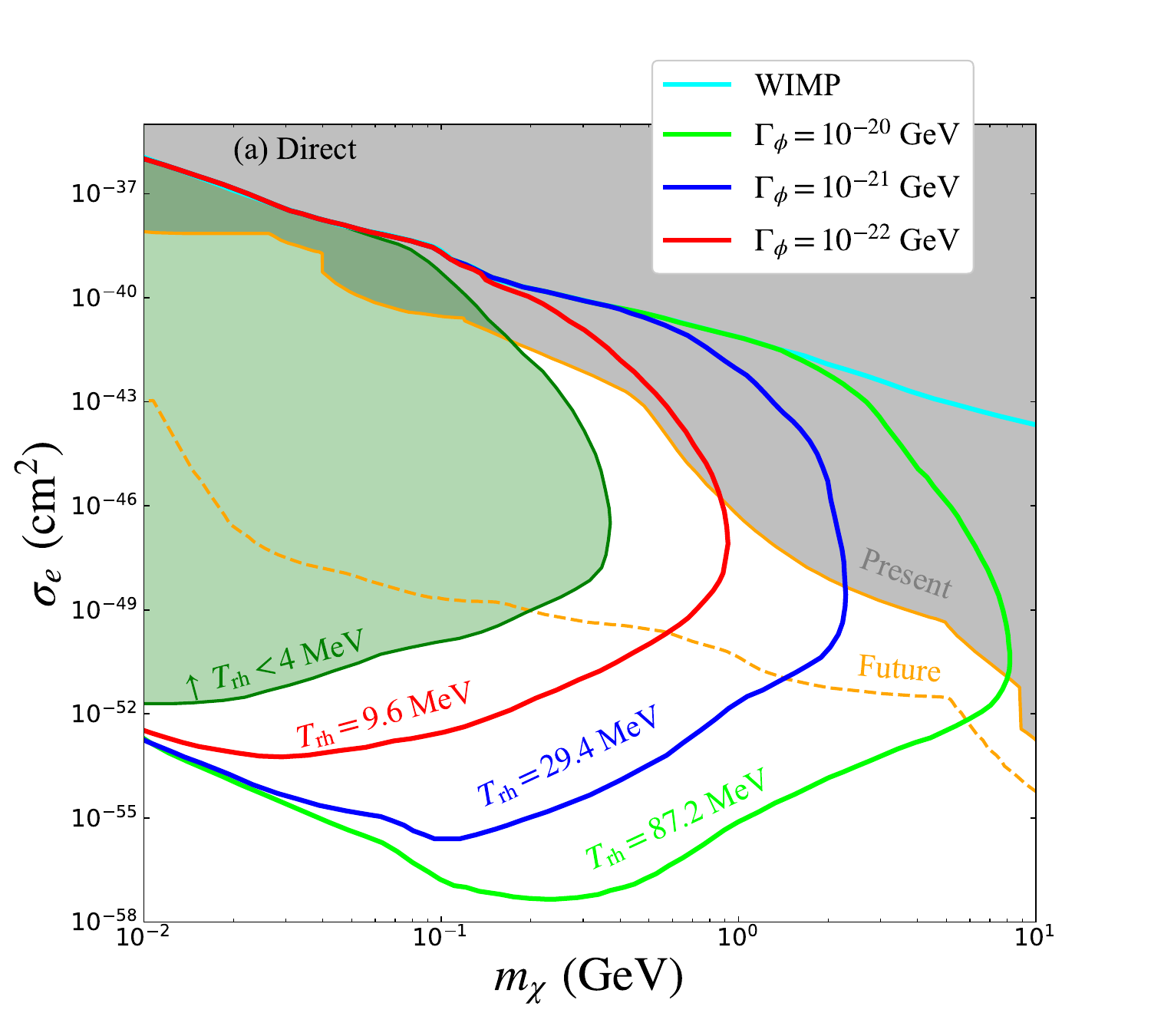}
		\includegraphics[width=0.45\linewidth]{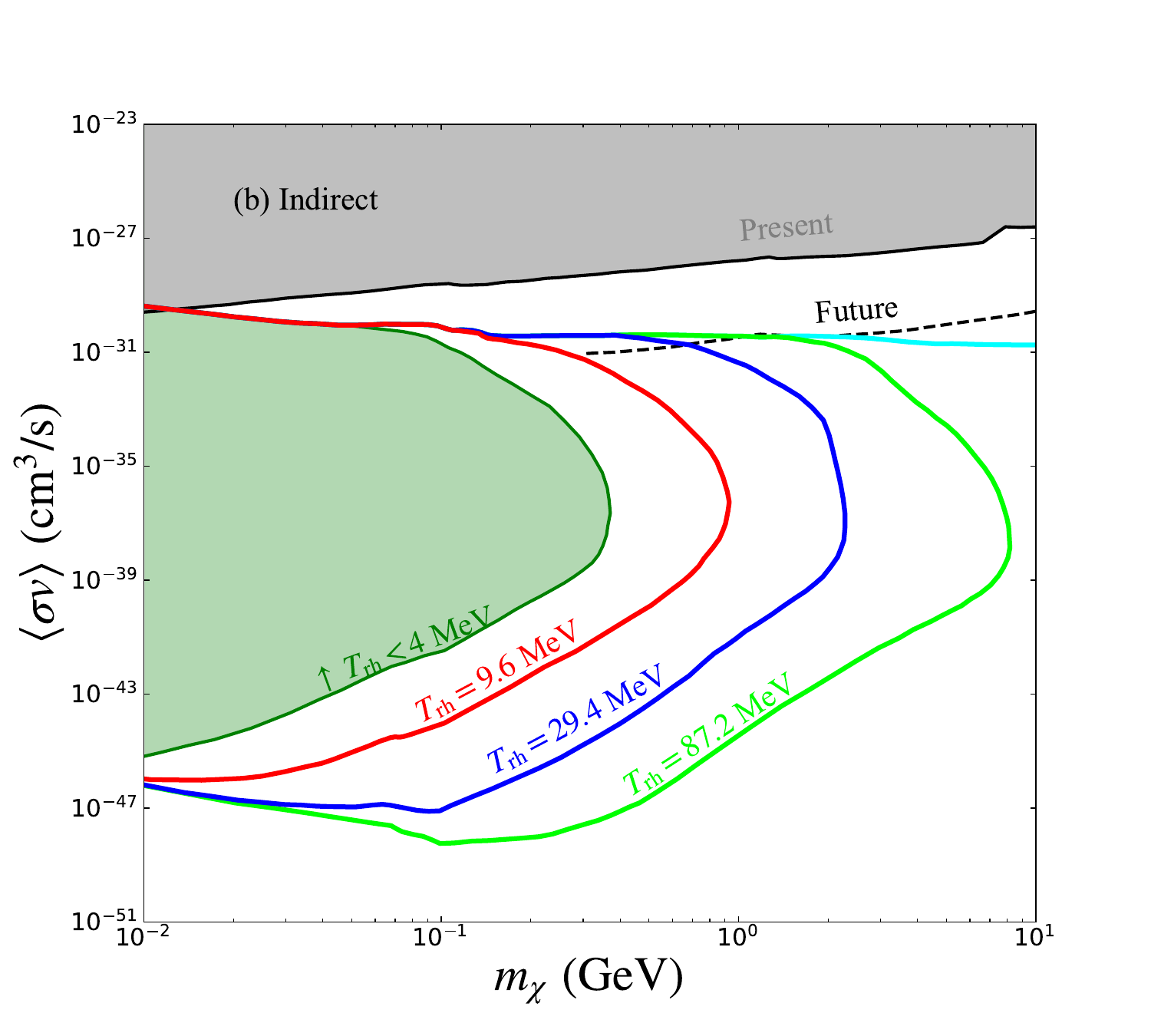}
		\includegraphics[width=0.45\linewidth]{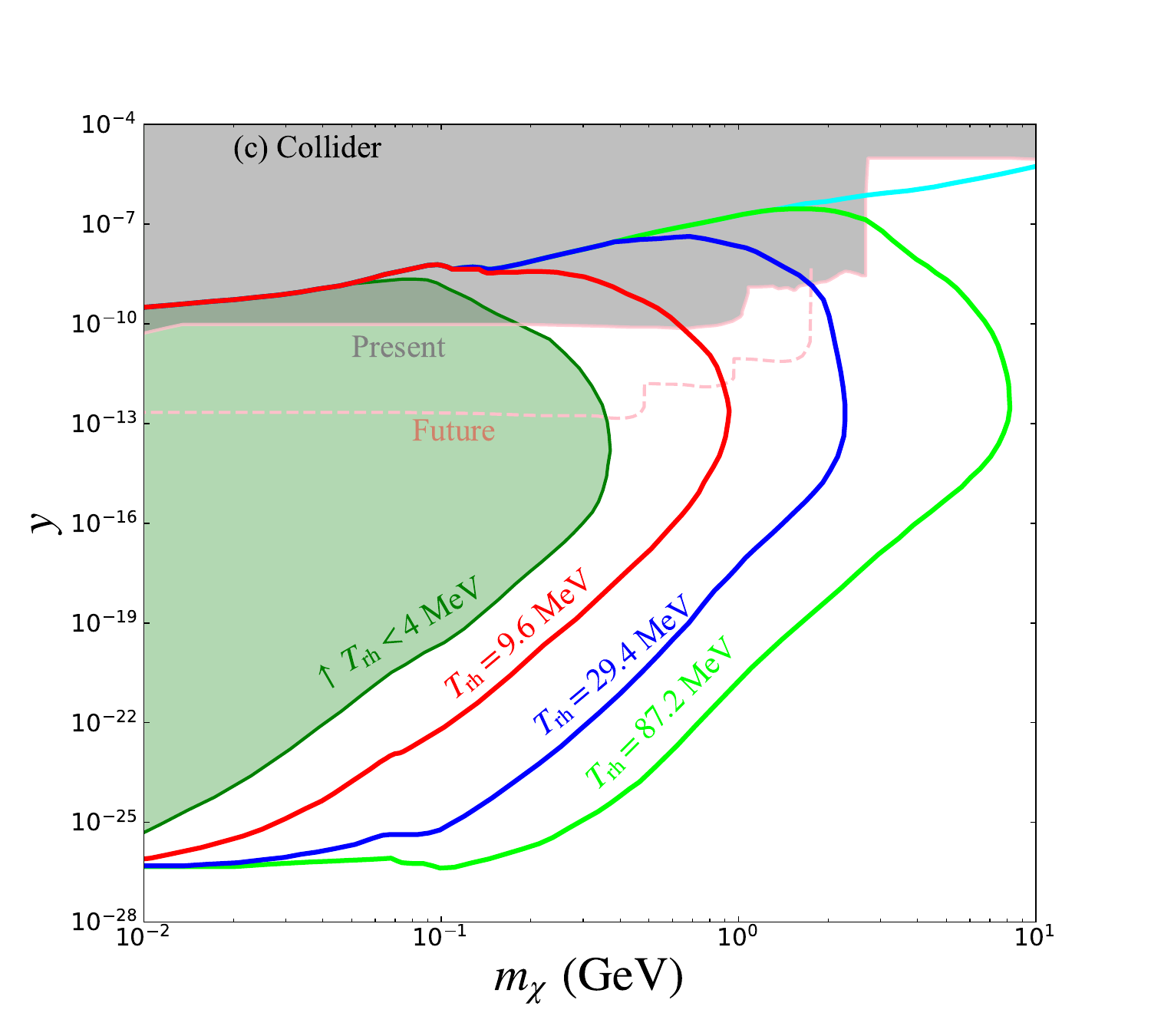}
		\includegraphics[width=0.45\linewidth]{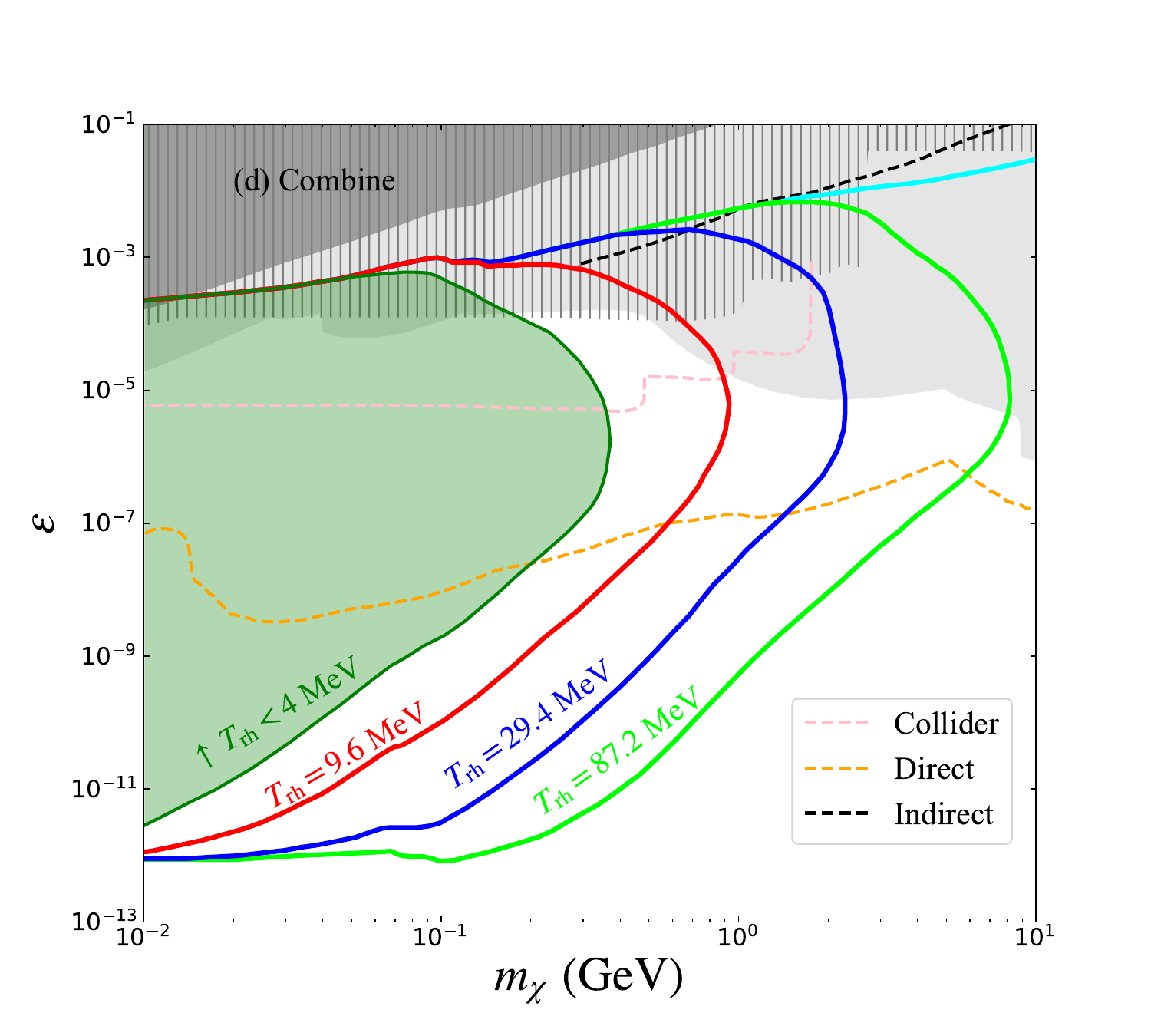}
	\end{center}
	\caption{Combined results for the complex scalar DM. The lime, blue, and red solid lines represent the three benchmark points that satisfy the observed dark matter relic density with different $\Gamma_\phi$ and corresponding $T_{\rh}$. The cyan solid line corresponds to the WIMP result. The green region indicates the disallowed $T_{\rh}$ by the BBN constraints\cite{Sarkar:1995dd, Kawasaki:2000en, Hannestad:2004px, DeBernardis:2008zz, deSalas:2015glj}. In panels (b), (c), and (d), all the benchmark lines have the same $\Gamma_\phi$ as in panel (a). The gray shaded regions denote the parameter space excluded by the current constraints, and the dashed lines indicate the sensitivities of future experiments. Panel (d) presents the combined constraints. For clarity, the light gray, dark gray, and line-shaded regions denote the constraints from the direct detection, indirect detection, and colliders, respectively. 
	}
	\label{FIG:fig2}
\end{figure}

In Figure~\ref{FIG:fig2} (a), the WIMP that matches the observed relic density is completely ruled out by direct detection limits.  However, including the low $T_{\rh}$ effect considerably improves the situation. The BBN bound excludes the region $T_{\rh}<4$ MeV, which translates into $\Gamma_\phi<1.8\times10^{-23}$ GeV via Equation~\eqref{Eqn:trh}.  Outside this region, when $T_{\rh}=9.6$ MeV with $\Gamma_\phi=10^{-22}$ GeV, the benchmark begins to show a clear departure from the WIMP for $m_\chi\gtrsim0.11$ GeV, and the required $\epsilon$  shift to smaller magnitudes as $m_\chi$ increases.  In the freeze-out regime, since $z_\chi$ cannot exceed the maximum value of $z_\chi^{\eq}$, there is an upper bound on $m_\chi$ \cite{Belanger:2024yoj}. Below this point of $(m_\chi\simeq0.88~\GeV, \sigma_e\simeq7.7\times10^{-49}~\cm^2)$, dark matter is produced via the freeze-in mechanism, so both $m_\chi$ and  $\epsilon$ decrease simultaneously until they reach a constant value. Using Equation~\eqref{Eqn:dd-e}, we get the minimum $\sigma_e\simeq5\times10^{-54}~\cm^2$ with $m_\chi\simeq0.03~\GeV$.  Naturally, increasing $T_\rh$ would allow for a larger $m_\chi$. For instance, when $T_{\rh}$ is raised to 87.2 MeV with $\Gamma_\phi=10^{-20}$ GeV, the deviation from WIMP starts at $m_\chi\simeq1.4~\GeV$. The maximum $m_\chi\simeq8$ GeV with $\sigma_e\simeq1.6\times10^{-51}~\cm^2$, and the minimum $\sigma_e\simeq4\times10^{-58}~\cm^2$ with $m_\chi\simeq0.25~\GeV$ are predicted. At the crossing point of the current direct detection constraints and BBN bound, the maximum allowed scattering cross section is $\sigma_e\simeq5.6\times10^{-42}~\cm^2$ with $m_\chi\simeq0.17~\GeV$.   The minimum $m_\chi$ that can be probed by future experiments is 0.17 GeV, and the testable $\sigma_e$ shrinks  with a growing $m_\chi$.

In the indirect detection constraints shown in panel (b) of Figure \ref{FIG:fig2},  due to the velocity suppression of the $p$‑wave annihilation for the scalar DM, almost all benchmark lines have $\langle \sigma v\rangle\lesssim \mathcal{O}(10^{-30})~\rm cm^3/s$, which lie below the current limits. The promising region $0.3~\GeV\lesssim m_\chi\lesssim1~\GeV$ with $\langle \sigma v\rangle\simeq10^{-31}~\rm cm^3/s$ is within the reach of future indirect experiments.
In panel (c) of Figure \ref{FIG:fig2}, the current collider constraints exclude the WIMP scenario for $m_\chi\lesssim2.7~\GeV$.  Within this mass range, dark matter with $T_{\rh}>4$ MeV is allowed when $y\lesssim\mathcal{O}(10^{-10})$. The future  STCF and DarkSHINE  can probe dark matter in the range of $0.18~\GeV\lesssim m_\chi\lesssim1.8~\GeV$ with $\mathcal{O}(10^{-13})\lesssim y \lesssim\mathcal{O}(10^{-10})$.

For the combined results in panel (d) of  Figure \ref{FIG:fig2}, it can be confirmed that the WIMP case with $m_{A'}=3m_\chi$ is completely ruled out. Among these three distinct kinds of experiments, the  direct detection provides the strongest limits both at present and in the future, which even rules out the promising parameter space of the future indirect detection experiments. The combined bounds determine that the allowed parameter space has the maximum $\epsilon\simeq1.3\times10^{-4}$ at $m_\chi\simeq0.18~\GeV$, and the minimum  $\epsilon\simeq10^{-12}$ at $m_\chi\simeq0.01~\GeV$. The future direct detection experiments will push the detection sensitivity  of dark photons  downward by nearly two orders of magnitude, which could probe the parameter space with $0.18~\GeV\lesssim m_\chi\lesssim10$ GeV and $\mathcal{O}(10^{-8})\lesssim \epsilon \lesssim\mathcal{O}(10^{-4})$. Meanwhile, the region $m_\chi\lesssim1~\GeV$ with $\epsilon\gtrsim5.5\times10^{-6}$ can additionally be checked with the future collider searches, which indicates $T_{\rh}$ less than 10 MeV.

\subsection{Dirac fermion scenario} \label{SUBSEC:DFS}
 
\begin{figure}
	\begin{center}
		\includegraphics[width=0.45\linewidth]{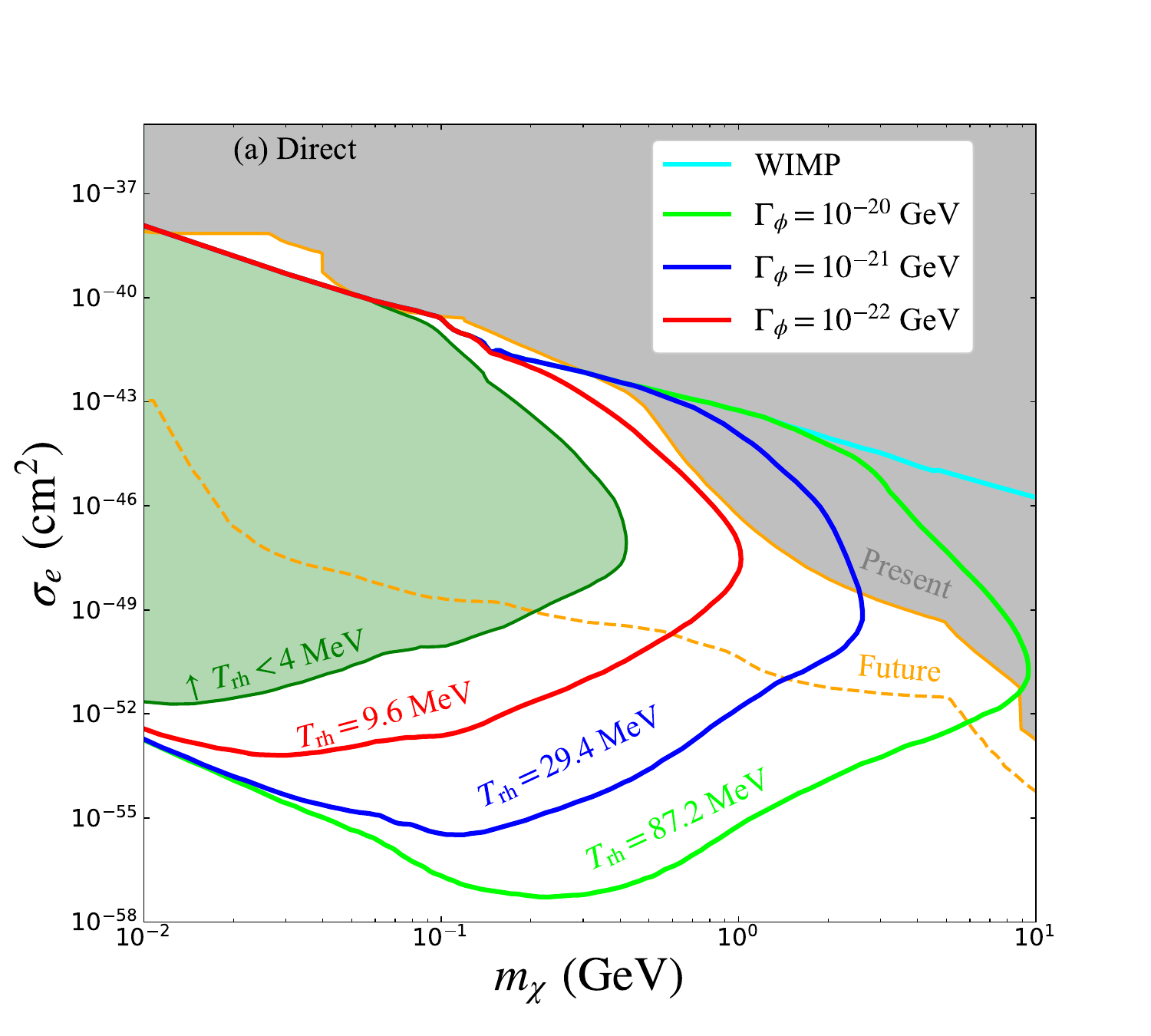}
		\includegraphics[width=0.45\linewidth]{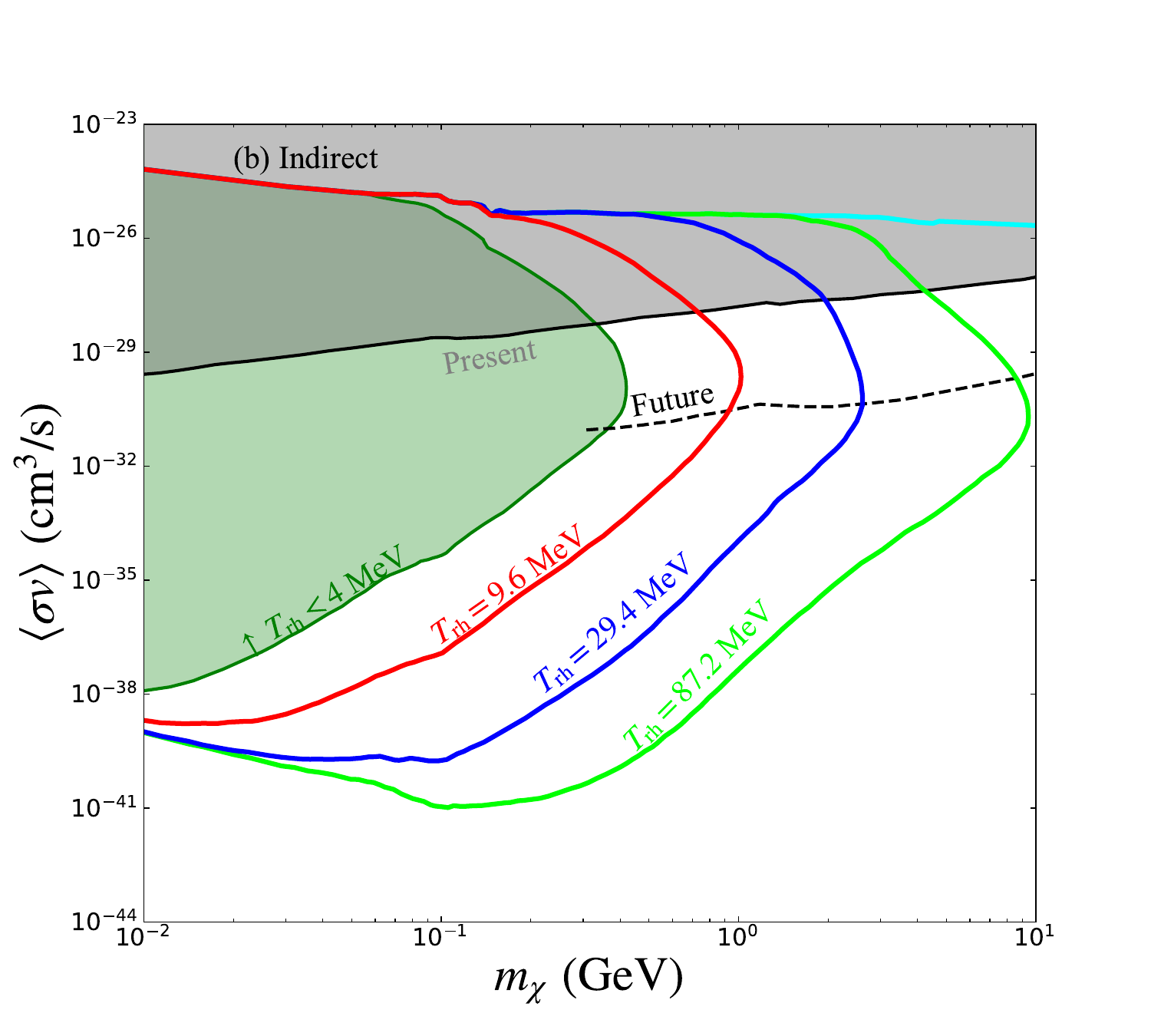}
		\includegraphics[width=0.45\linewidth]{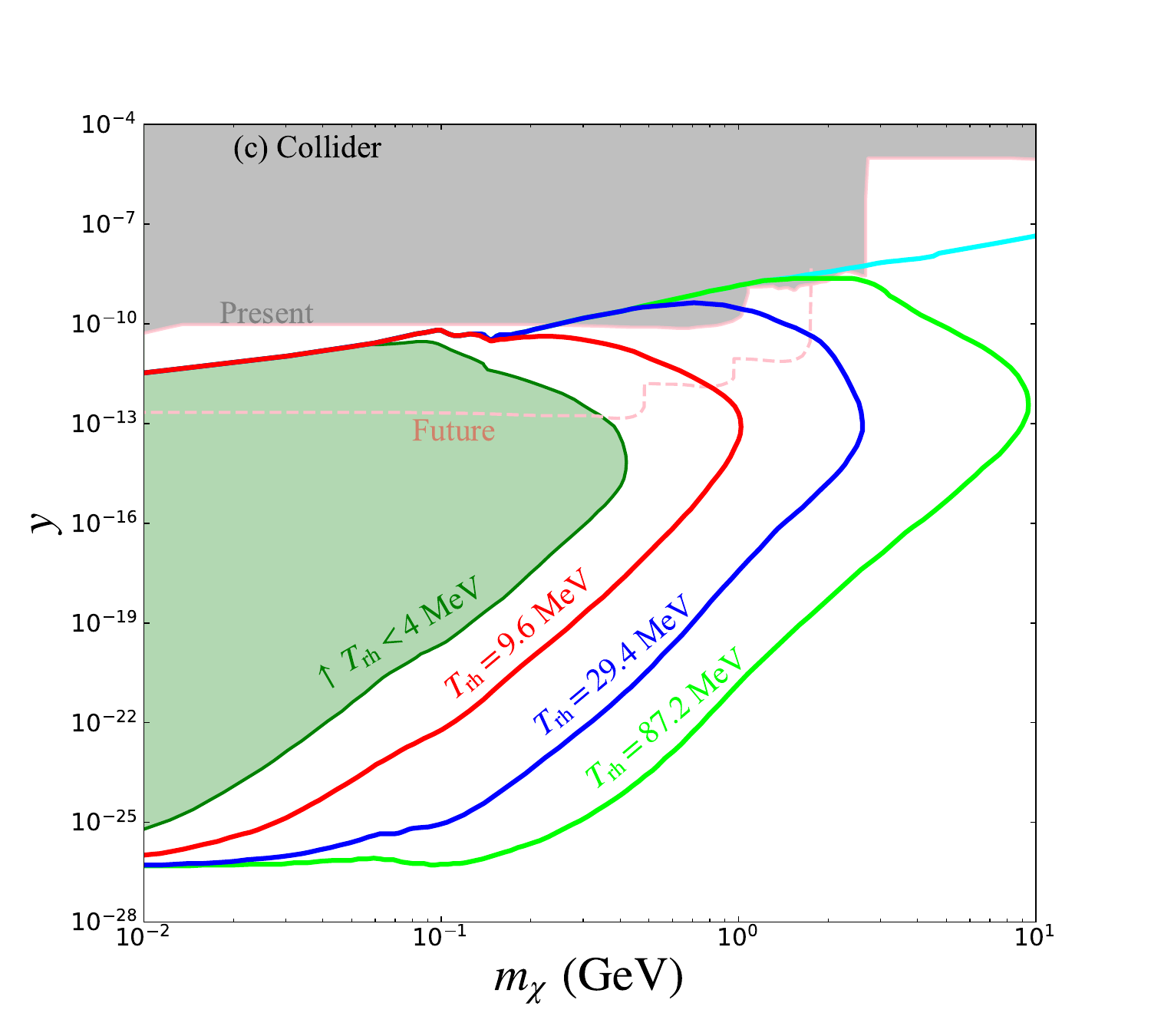}
		\includegraphics[width=0.45\linewidth]{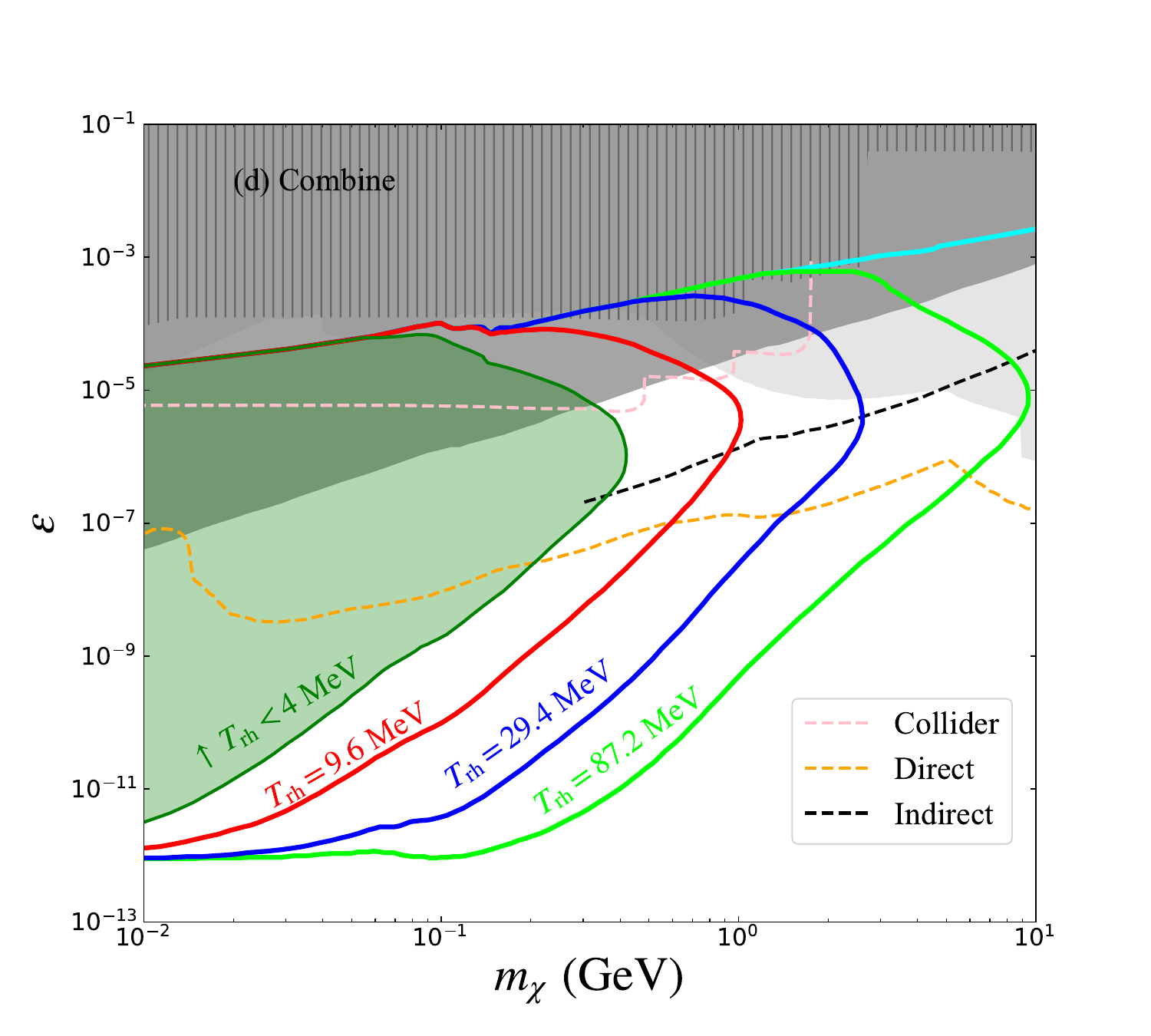}
	\end{center}
	\caption{Same as Figure~\ref{FIG:fig2} but for Dirac fermion DM.
	}
	\label{FIG:fig3}
\end{figure}

The direct detection constraints of Dirac fermion DM are shown in panel (a) of Figure~\ref{FIG:fig3}.  Compared with the scalar scenario, the Dirac WIMP benchmark has a slightly lower scattering cross section $\sigma_e$. The variation behavior of the benchmark lines with different $T_{\rh}$ is similar to that in the scalar case. Beyond the BBN bound, as $T_{\rh}$ increases to 87.2 MeV, the maximum attainable $m_\chi$  grows from 0.4 GeV with $\sigma_e\simeq3.6\times10^{-49}~\cm^2$ to 9.6 GeV with $\sigma_e\simeq10^{-51}~\cm^2$. And this benchmark has the minimum $\sigma_e\simeq5\times10^{-58}~\cm^2$ when $m_\chi\simeq0.22$ GeV.  Both the upper and lower boundaries have decreasing $\sigma_e$  with increasing $T_{\rh}$. The future experiments will be sensitive to the region with $\mathcal{O}(10^{-55})~\cm^2 \lesssim \sigma_e \lesssim \mathcal{O}(10^{-41})~\cm^2$.

In panel (b) of Figure~\ref{FIG:fig3}, the $s$‑wave processes  with relatively  large $\langle \sigma v\rangle$ lead to the exclusion of WIMP  by current indirect detection limits.  The upper boundary for the allowed parameter space of $\langle \sigma v\rangle$ is determined by the present constraints. The required maximum $\langle \sigma v\rangle$ rises from $5\times10^{-29}~\rm cm^3/s$ to $10^{-27}~\rm cm^3/s$ when $m_\chi$ increasing from sub-GeV to 10 GeV. Future experiments will be able to probe $0.36~\GeV\lesssim m_\chi\lesssim10$~GeV within $\mathcal{O}(10^{-31})~\cm^3/s \lesssim \langle \sigma v\rangle \lesssim \mathcal{O}(10^{-28})~\cm^3/s$.  

In panel (c) of Figure~\ref{FIG:fig3}, within the collider-sensitive range , $m_\chi$  less than 0.2 GeV or larger than 2.7 GeV is not excluded by current collider constraints for the WIMP benchmark.  The benchmarks with $T_{\rh}>4$ MeV  satisfy the present bounds in the region of $y\lesssim\mathcal{O}(10^{-10})$.  The promising dark matter  has $0.07~\GeV\lesssim m_\chi\lesssim1.8~\GeV$ with $\mathcal{O}(10^{-13})\lesssim y \lesssim\mathcal{O}(10^{-10})$.

In panel (d) of Figure~\ref{FIG:fig3}, we report that the WIMP scenario is still excluded under the combined constraints.  For $\epsilon\lesssim2.5\times10^{-5}$, the indirect detection and BBN provide the most stringent limits when  $m_\chi\lesssim0.8~\GeV$, whereas in the opposite mass range, the direct detection has stronger exclusion ability.  Among the future experiments, the direct detection offers the best sensitivity, which would probe  $0.2~\GeV\lesssim m_\chi\lesssim10~\GeV$ with $\mathcal{O}(10^{-8})\lesssim \epsilon \lesssim\mathcal{O}(10^{-5})$. This broadens the detection range of dark photons compared with future collider sensitivity,  which is hardly allowed by the current combined constraints. A narrower parameter range, e.g., $m_\chi\lesssim5$ GeV with $\epsilon\gtrsim\mathcal{O}(10^{-7})$, is promise to be probed simultaneously by direct and indirect experiments, but the associated $T_{\rh}$ is required to be smaller than 60 MeV.

\subsection{Majorana fermion scenario} \label{SUBSEC:MFS}

\begin{figure}
	\begin{center}
		\includegraphics[width=0.45\linewidth]{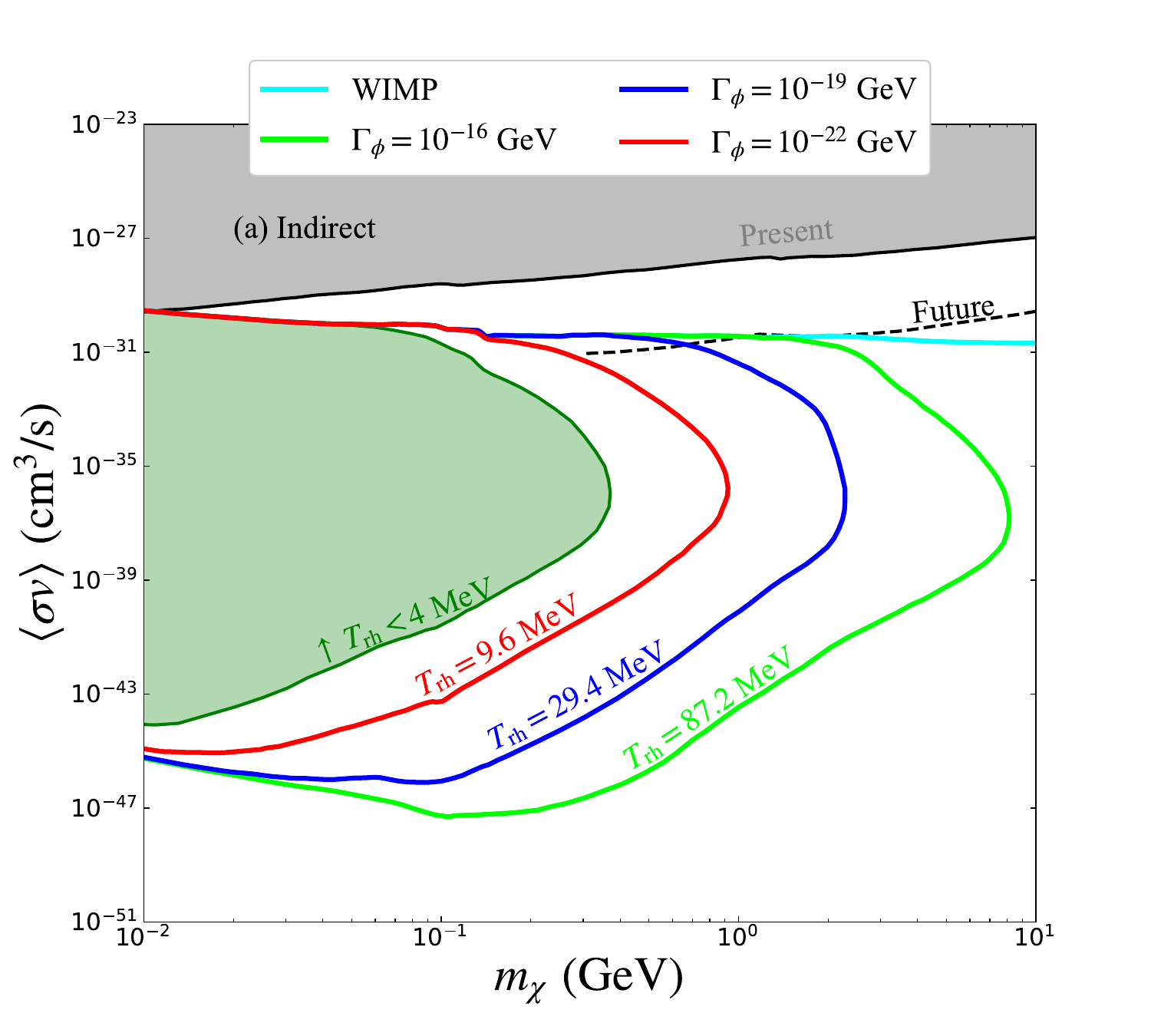}
		\includegraphics[width=0.45\linewidth]{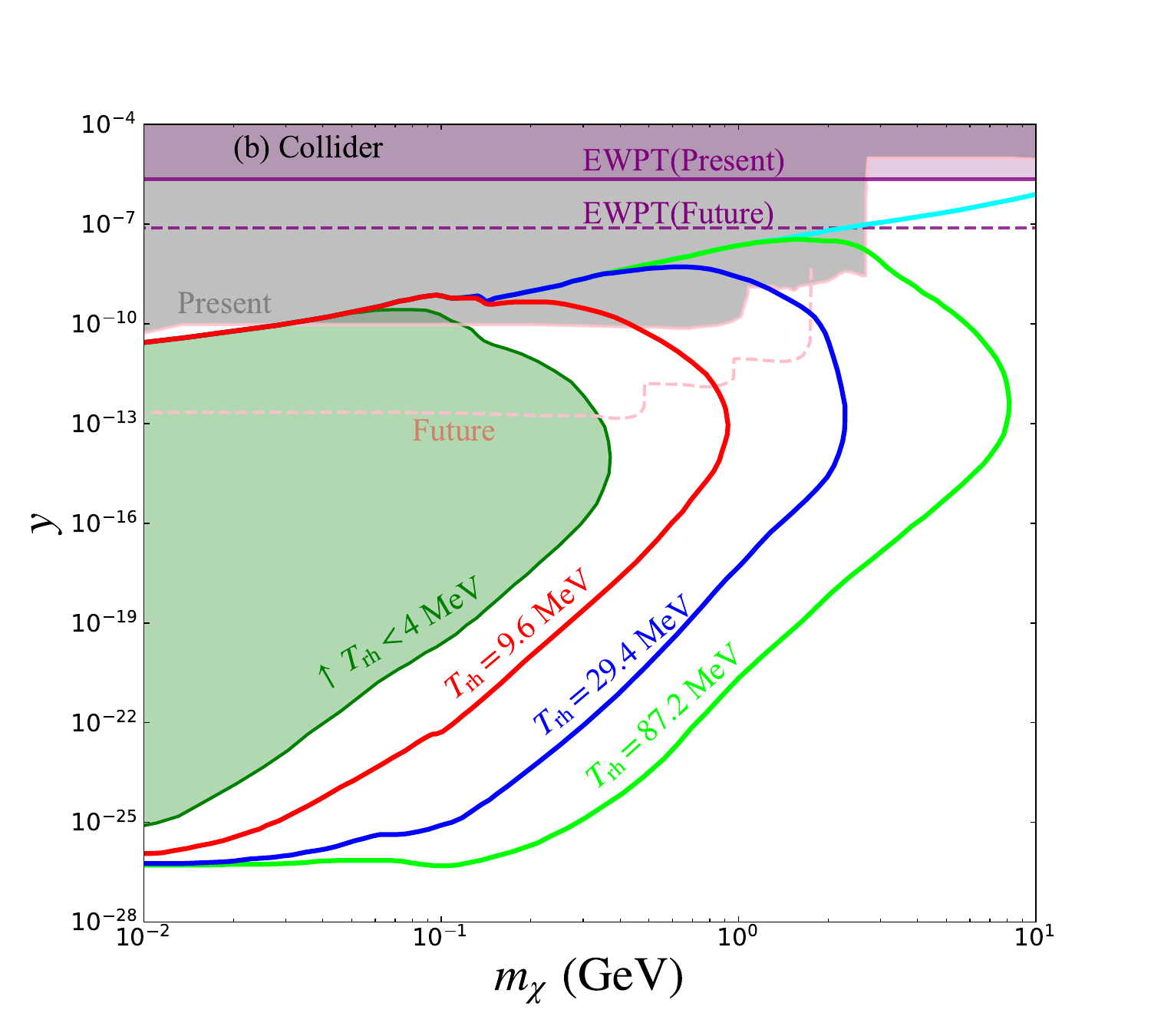}
		\includegraphics[width=0.45\linewidth]{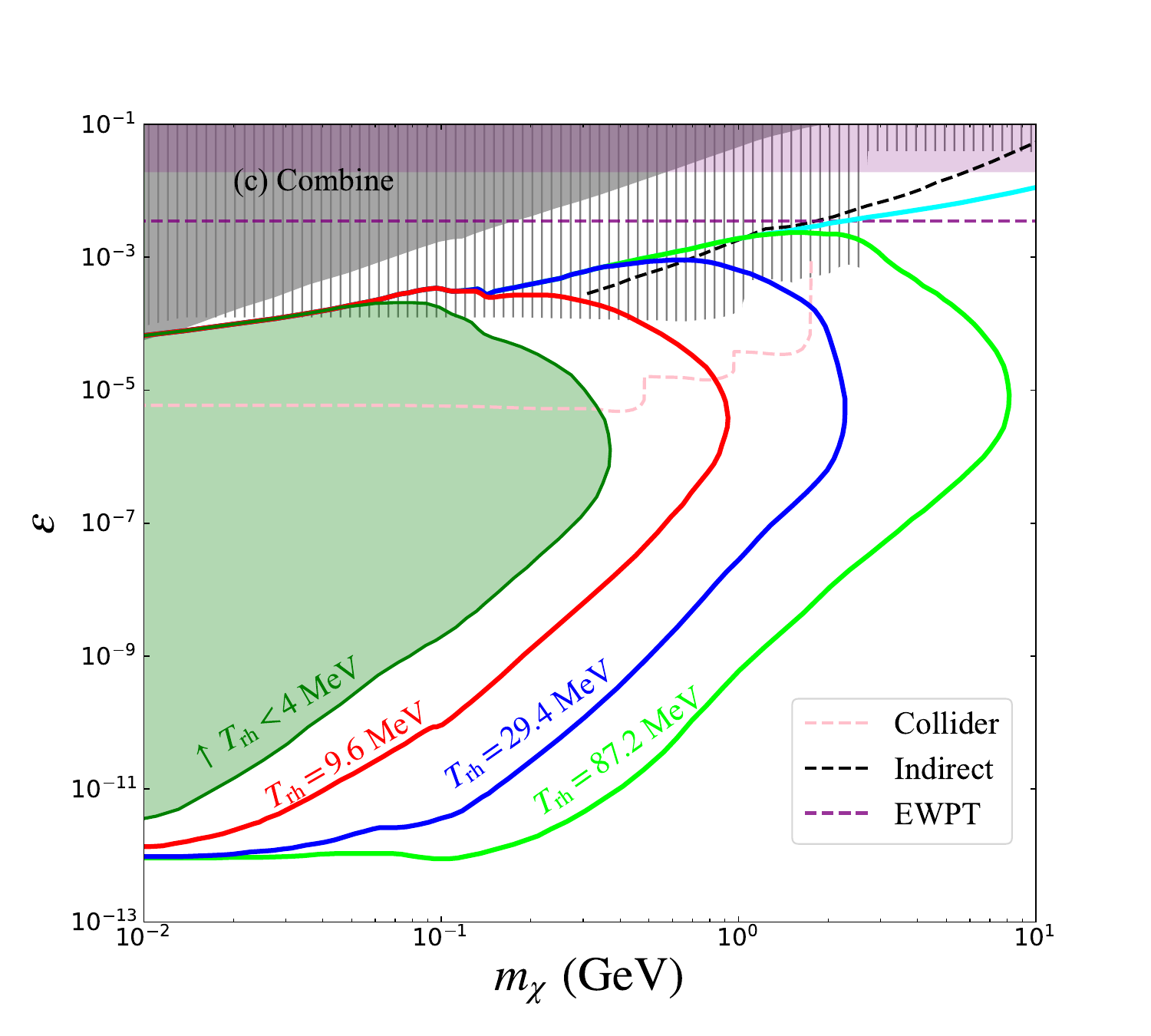}
	\end{center}
	\caption{Same as Figure~\ref{FIG:fig2} but for Majorana fermion DM. Notably, as we mentioned in Subsection \ref{SUBSEC:DD}, none of the various direct detection constraints are applicable to this scenario, therefore, we do not show the direct detection bounds. In panels (b) and (c), the purple and the solid lines respectively indicate the current exclusion  and  the projected sensitivity  from electroweak precision tests of $Z$ boson.
	}
	\label{FIG:fig4}
\end{figure}

Panel (a) of Figure~\ref{FIG:fig4} corresponds to the  indirect detection constraints. The dark matter annihilation processes are also $p$‑wave dominant in the Majorana fermion scenario, thus  $\langle \sigma v\rangle$  are  suppressed by the velocity of DM. In this case, all the benchmarks have $\langle \sigma v\rangle\lesssim \mathcal{O}(10^{-30})~\rm cm^3/s$.  In the region not affected by BBN, as $T_{\rh}$ increases to 87.2 MeV, the maximum  attainable $m_\chi$ of the benchmarks rises from 0.37 GeV to 8.3 GeV, and the minimum $\langle \sigma v\rangle$ decreases from $8\times10^{-45}~\rm cm^3/s$ to $4.7\times10^{-48}~\rm cm^3/s$.  The predicted upper limit on $\langle \sigma v\rangle$ is determined by the WIMP case. The current experimental results do not pose any threat to the benchmarks including the WIMP.  Dark matter in the region of $0.3~\GeV\lesssim m_\chi\lesssim1~\GeV$ with $\langle \sigma v\rangle\simeq10^{-31}~\rm cm^3/s$ lies within the reach of future experimental searches.

In panel (b) of Figure~\ref{FIG:fig4}, the currently collider safe region  $\mathcal{O}(10^{-13})\lesssim y\lesssim\mathcal{O}(10^{-10})$ with $0.11~\GeV\lesssim m_\chi\lesssim1.8~\GeV$  is promising for future experimental tests, provided that the required $T_{\rh}$ is below 60 MeV.  For $m_\chi\gtrsim2.7~\GeV$, the collider constraints are fairly loose. The current EWPT bounds exclude  $y \lesssim2\times10^{-6}$. The future sensitivity will be one order of magnitude lower, which can probe the WIMP with $m_\chi\lesssim10~\GeV$ as well as  the low-temperature reheating DM with $T_{\rh}\sim100$~MeV.

Finally, for the combined constraints shown in panel (c) of Figure~\ref{FIG:fig4}, due to the lack of direct detection constraints, the collider bound becomes the strongest one  when $m_\chi\lesssim2.7~\GeV$. Within this mass range, the WIMP  and the expected sensitivity of future indirect detection both fall into a dangerous situation. Outside this mass range, the EWPT constraint disallows $\epsilon\gtrsim0.019$, while the WIMP with $m_\chi\lesssim10~\GeV$ is not affected by this bound.  The prospect for probing dark matter relies mainly on upcoming collider experiments, which are sensitive to $0.1~\GeV\lesssim m_\chi\lesssim1.8~\GeV$ with $\epsilon \sim\mathcal{O}(10^{-5})$. The future EWPT outcome  will probe GeV scale DM.

\section{Conclusion} \label{SEC:CL}

Since the conventional non‑resonant dark photon $A^\prime$ portal dark matter can hardly survive under the stringent constraints from direct detection, indirect detection, and collider searches, we investigate the dark matter production affected by a low reheating temperature $T_{\rh}$, which is created by the delayed decay of the inflaton $\phi$ into SM radiation. We discuss the improvement of this low-temperature reheating scenario relative to the traditional WIMP case under the combined constraints.  The non-resonance benchmark set $m_{A^\prime}/m_\chi=3$ and $\alpha_\chi=0.5$ is considered. The analysis is elaborated in three distinct scenarios, namely, complex scalar DM, Dirac fermion DM, and Majorana fermion DM.

In the complex scalar scenario, the WIMP case is completely ruled out by the present combined constraints. Dark matter with a low reheating temperature $4~\MeV \lesssim T_{\rh}\lesssim\mathcal{O}(100)~\MeV$ survives when $m_\chi\lesssim10$ GeV and $\epsilon\lesssim\mathcal{O}(10^{-4})$. The projected sensitivity of future direct detection experiments is able to probe $0.18~\GeV\lesssim m_\chi\lesssim10~\GeV$ and $\mathcal{O}(10^{-8})\lesssim \epsilon \lesssim\mathcal{O}(10^{-4})$,  which significantly enlarges the parameter space for dark photon searches relative to colliders.

In the Dirac fermion scenario, WIMP is still excluded by the combined constraints. Due to the $s$-wave annihilation of dark matter,  the indirect detection constraint stands out, which leads to a certain shrinkage of the allowed parameter region with $\epsilon\lesssim2.5\times10^{-5}$ relative to the complex scalar case. Future direct detection will be able to probe $0.2~\GeV\lesssim m_\chi\lesssim10~\GeV$ with $\mathcal{O}(10^{-8})\lesssim \epsilon \lesssim\mathcal{O}(10^{-5})$. The required $T_{\rh}$ is roughly the same as that in the scalar scenario. In contrast, the future indirect detection experiments are sensitive to only a more restricted parameter space, and the projected collider sensitivities  are almost  excluded by the present constraints.

The Majorana fermion scenario is incompatible with the direct detection constraints, while the indirect detection limits are quite weak. As a result, collider constraints become the most stringent one for light $m_\chi$, which could exclude $m_\chi\lesssim2.7~\GeV$ with $\epsilon\gtrsim10^{-4}$. The future colliders will be sensitive to $0.1~\GeV\lesssim m_\chi\lesssim1.8~\GeV$ with $\epsilon \sim\mathcal{O}(10^{-5})$. For larger $m_\chi$, the EWPT constraints become effective for $\epsilon\gtrsim3.5\times10^{-3}$, and can only probe a very small portion of the benchmarks.

In summary, the WIMP dark matter benchmark $m_{A'}/m_\chi=3$ within $m_\chi\lesssim10~\GeV$ is nearly ruled out in all three scenarios, whereas dark matter under the influence of a low $T_{\rh}$ can avoid the present constraints. The  complex scalar and  Dirac fermion scenarios primarily depend  on the future direct detection experiments to probe $\chi$ and $A^\prime$, while the Majorana fermion scenario  relies more on the collider searches, whose accessible parameter region is far smaller than that in the first  two cases.

\section*{Acknowledgments}

This work is supported by the National Natural Science Foundation of China under Grant  No. 12275134 and No. 12335005, Natural Science Foundation of Shandong Province under Grant No. ZR2026QC0016 and No. ZR2026MS0091, and University of Jinan Disciplinary Cross-Convergence Construction Project 2024 (XKJC-202404).

%%%%%%%%%%%%%%%%%%%%%%%%%%%%%

\end{document}